\documentclass[prd, aps, superscriptaddress, preprintnumbers, twocolumn, floatfix, nofootinbib]{revtex4-2}
\pdfoutput=1

\usepackage{amsfonts}
\usepackage{amsmath}
\usepackage{amssymb}
\usepackage{bm}
\usepackage{cancel}
\usepackage{dcolumn}
\usepackage{graphicx}   
\usepackage[latin1]{inputenc}
\usepackage{latexsym}
\usepackage{rotating}
\usepackage{hyperref}
\usepackage{graphicx}
\usepackage{ulem}
\usepackage{xcolor}
\usepackage{cases}

\newcommand\be{\begin{equation}}
\newcommand\ba{\begin{eqnarray}}
\newcommand\ee{\end{equation}}
\newcommand\ea{\end{eqnarray}}

\newcommand{\Msol}{\ensuremath{M_{\odot}}}
\newcommand{\nn}{\nonumber \\}
\newcommand{\gsim}{\mathrel{\hbox{\rlap{\lower.55ex \hbox {$\sim$}}
                   \kern-.3em \raise.4ex \hbox{$>$}}}}
\newcommand{\lsim}{\mathrel{\hbox{\rlap{\lower.55ex \hbox {$\sim$}}
                   \kern-.3em \raise.4ex \hbox{$<$}}}}

\newcommand{\tbf}{\textbf}

\begin{document}

\title {Direct Collapse Supermassive Black Holes from Ultralight Dark Matter}

\author{Hao Jiao}
\email{hao.jiao@mail.mcgill.ca}
\affiliation{Department of Physics, McGill University, Montr\'{e}al, QC, H3A 2T8, Canada} 

\author{Robert Brandenberger}
\email{rhb@physics.mcgill.ca}
\affiliation{Department of Physics, McGill University, Montr\'{e}al, QC, H3A 2T8, Canada} 

\author{Vahid Kamali}
\email{vahid.kamali2@mcgill.ca}
\affiliation{Department of Physics, McGill University, Montr\'{e}al, QC, H3A 2T8, Canada}
\affiliation{Department of Physics, Bu-Ali Sina (Avicenna) University, Hamedan 65178,  016016, Iran}
\affiliation{School of Continuing Studies, McGill University, Montr\'{e}al, QC, H3A 2T5, Canada}
\affiliation{Trottier Space Institute, Department of Physics, McGill University, Montr\'{e}al, QC, H3A 2T8, Canada}

\date{\today}


\begin{abstract}

We study the possibility that parametric resonant excitation of photons in an ultralight dark matter halo could generate the required flux of Lyman-Werner photons to allow the direct collapse formation of supermassive black hole seeds.   

\end{abstract}

\maketitle

\section{Introduction} 
\label{sec:intro}
 
The origin of supermassive black holes (SMBHs) remains a mystery, in particular, the origin of such black holes with mass $\gsim 10^9 M_\odot$ at high redshifts (see e.g. \cite{Marta} for a review).  
An important channel of SMBH seed formation is via the {\it direct collapse} of atomic cooling gas clouds, 
which will lead to the formation of massive black holes with initial masses about $10^5 M_\odot$ \cite{DCBH-1,DCBH-2,DCBH-3,DCBH-4,DCBH-5,DCBH-6,DCBH-7,DCBH-8,DCBH-9,DCBH-10,DCBH-11}. Such ``heavy'' seeds\footnote{The ``heavy'' seeds of SMBH are in contrast to the ``light'' seeds of $10^3M_\odot$ from the remnants of Pop III stars.} are able to grow to $10^9M_\odot$ at redshifts $z>6$ without super-Eddington accretion.

In order to form a direct collapse black hole (DCBH), some criteria need to be satisfied, the most important one of which is a sufficient flux of ultraviolet (UV) photons. Strong radiation in the energy band $0.76eV < E < 13.6eV$ will suppress the formation of molecular hydrogens via photodissociation and photodetachment. The existence of $H_2$ would cool the gas cloud to significantly below the atomic cooling threshold, which would lead to the fragmentation of the cloud and would prevent the direct collapse. Additionally, this UV flux needs to be present before the epoch of star formation since metals produced by the stars will also cool the cloud. New physics is required in order to produce this flux.

In \cite{Bryce} it was proposed that electromagnetic radiation from a scaling distribution of superconducting cosmic string loops (see e.g. \cite{CSrevs} for reviews of cosmic strings) could provide the required radiation.  In this case,  the cosmic strings also provide an additional source of high redshift nonlinear seed fluctuations which would enable DCBH formation even at significantly higher redshifts than the ones for which we currently have data.  A scaling distribution of superconducting string loops is characterized by the two parameters $\mu$ and $I$,  where $\mu$ is the string tension and $I$ is the current. A small region in parameter space was identified in which the direct collapse scenario can be realized \footnote{The nonlinear halo mass function in a model with a component of cosmic string loops superimposed on a spectrum of primordial adiabatic $\Lambda$CDM fluctuations has recently been worked out analytically in \cite{JH1} and the results were confirmed with N-body \cite{JH2} and hydrodynamical \cite{JH3} simulations (see also \cite{Bram} and \cite{IMBH} for earlier work)}. 

In a recent paper \cite{Kusenko}, the possibility was explored that relic particle decay (e.g. decaying dark matter) could provide the conditions for DCBH formation. Here we explore a different channel.  We assume that the dark matter is made up of ultra-light pseudo-scalar ``particles'' -- wave dark matter (see \cite{Elisa, Hui} for reviews of wave dark matter).  The corresponding fields are typically coupled to electromagnetism via a Chern-Simons term.  In a dark matter halo, the coherent oscillation of the dark matter field (which we will call ``axion'' in the following) will induce parametric resonance of infrared photons in a similar way that the oscillations of the inflaton field after the end of a period of primordial inflation will induce a parametric resonant instability for all fluctuations coupled to the inflaton (see \cite{DK, TB} for the original work, \cite{KLS1, STB, KLS2} for later studies, and \cite{RHrevs} for reviews).  We explore the idea that these photons could provide the required flux of Lyman-Werner photons or UV photons. We identify the parameter space in which this mechanism can be realized assuming either that the photons thermalize in the halo (which can occur since the halos are optically thick at high redshifts), or that there is a sufficiently strong cascade of the photons generated via the parametric resonance instability. \footnote{Note that our mechanism allows supermassive black hole formation via the direct black hole collapse mechanism even if the cosmological fluctuations are Gaussian as in the standard cosmological paradigm. On the other hand, non-Gaussian seeds, such as cosmic strings, could facilitate this mechanism.} 

In the following section, we review the evolution of axions with a Chern-Simons term. Section III discusses the parametric resonance of the electromagnetic gauge field. The DCBH formation criteria and how our model realizes these criteria are presented in Sections IV and V, respectively. Finally, we conclude with a discussion of this model in Section VI.

\section{Ultralight Axion Dark Matter Distribution inside of a Galactic Halo}

We consider dark matter to be described by an ultralight dark matter field $\phi$ with a generic potential which can be expanded about its minimum as
\be
V(\phi) \, = \, \frac{1}{2} m^2 \phi^2 \, ,
\ee
where $m$ is the mass of $\phi$.  Cosmological considerations lead to a lower bound \cite{lower} on $m$ of the order $m > 10^{-20} {\rm{eV}}$ (for lower values of the mass, structure formation on scales measured by the Lyman $\alpha$ forest would be suppressed) while demanding that the dark matter is wavelike leads to the upper bound $m < 10 {\rm{eV}}$ (see e.g. \cite{Rodd} for a recent discussion). 

If $\phi$ is an axion-like particle and hence a pseudo-scalar, then it is rather generic to assume that there is a Chern-Simons type coupling to the gauge field $A_{\mu}$ of electromagnetism. Thus, we consider the Lagrangian
\be
{\cal{L}} \, = \, -\frac12 (\partial\phi)^2 + V(\phi) -\frac14 F_{\mu\nu}F^{\mu\nu} + g_{\phi\gamma}\phi F_{\mu\nu}\tilde{F}^{\mu\nu},
\ee
where $F_{\mu \nu}$ is the field strength tensor of $A_{\mu}$ and ${\tilde{F}_{\mu \nu}}$ is its dual.  The coupling constant $g_{\phi \gamma}$ has inverse mass units.  From the non-observation of events coming from interactions of dark matter with photons, there are mass-dependent upper limits on the coupling $g_{\phi \gamma}$ which are shown in Fig. 1 (taken from \cite{AxionLimits}) \footnote{See \cite{constraint} for an older review.}.

\begin{figure}
  	\includegraphics[width=8cm]{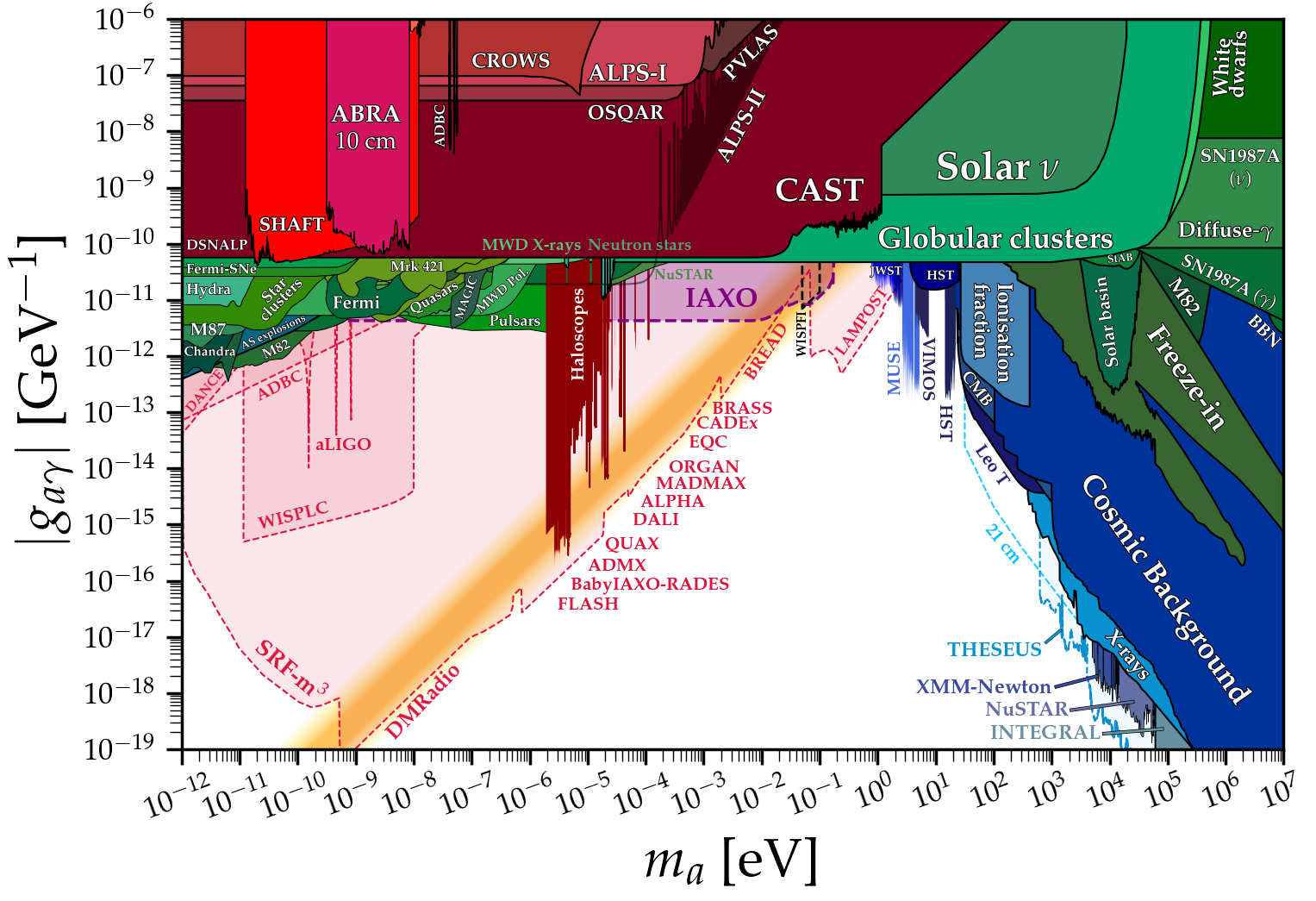}
	\caption{Summary of the bounds on the coupling $g_{\phi \gamma}$ as a function of the mass $m_{\phi}$ taken from \cite{AxionLimits}. Solid curves correspond to current constraints while dashed curves are projections.}
	\label{fig0}
\end{figure}

Note that the Chern-Simons coupling violates the CP symmetry since it affects the two-photon polarization states differently. Hence, this term allows the generation of primordial electromagnetic fields, a mechanism which has been known for a long time (see e.g. \cite{JF1, JF2, Joyce, Juerg}).

We will now consider the evolution of the dark matter field inside of a gravitationally bound halo. In this case, we can neglect the expansion of space, and the equation of motion for $\phi$ (in the absence of back-reaction by the gauge field) is
\be
- \partial_t^2\phi + \frac{1}{r^2}\partial_r \big(r^2\partial_r\phi\big) - m^2\phi \, = \, 0 \ ,
\ee
where $r$ is the radial coordinate measured from the center of the halo.  Approximating $\phi$ to be homogeneous within the halo we obtain an oscillatory solution
\be \label{ansatz}
\phi(t) = \phi_0 \sin(mt + \alpha) \, ,
\ee
where $\alpha$ is a phase determined by the initial conditions. If we assume that $\phi$ makes up all of the dark matter, we can obtain the amplitude $\phi_0$ by matching the energy density in the halo with the scalar field energy density
\be
\rho \, = \, \frac12 m^2\phi^2(t),
\ee
A nontrivial dark matter density profile $\rho_{DM}(r)$ can be taken into account by allowing $\phi_0$ to depend on $r$.

Let us now estimate the value of $m \phi_0$ which is relevant to our study of direct collapse black hole formation.  Let us consider an approximately spherical halo which virializes at a redshift $1 + z = 20$. The overall energy density $\rho_h$ inside of the halo is then
\be
\rho_h \, \sim \, \Delta_{vir} \rho_c (1+z)^3 ,
\ee
where $\rho_c$ is the current average background density, $\Delta_{vir}\sim200$ is the ratio between virialized halo density and background density at the time of formation \cite{Zeld}, and the factor $(1+z)^3$ is due to the redshifting of the background density.  Inserting the value of $\rho_c$ we obtain
\be 
\rho_h \, \sim \, 5\times 10^{-40} {\rm{GeV}}^4 \, , \label{eq-halo-dens}
\ee
which leads to the result
\be \label{rel1}
m \phi_0 \, \sim  \, 3\times 10^{-20} GeV^2.
\ee

In light of the constraint from Fig. 1, we will parametrize the coupling constant $g_{\phi \gamma}$ as
\be
g_{\phi \gamma} \, \equiv {\tilde g}_{\phi \gamma} 10^{-10} {\rm GeV}^{-1} 
\ee
with the constraint ${\tilde g}_{\phi \gamma} \ll 1$.

\section{Resonant Production of Photons from an Oscillating Dark Matter Field}

Let us now consider the equation of motion for the electromagnetic gauge field $A_{\mu}$ in the presence of the oscillating ultralight dark matter field $\phi$ inside of the halo.\footnote{As discussed in \cite{Juerg}, the effects of the residual free electrons is negligible.}
\be
\ddot A_k + \Big[ k^2 \pm 4 g_{\phi\gamma} k\dot\phi \Big]A_k \, = \, 0 \, ,
\ee
where the $\pm$ apply to the different polarization states of the photon.  Inserting the ansatz (\ref{ansatz}) for $\phi$, setting the phase $\alpha$ to zero, and expressing the result in terms of the dimensionless temporal variable ${\tilde{t}} \equiv mt$ we obtain
\be 
\frac{d^2 A_k}{d{\tilde{t}}^2} + \bigg[ \frac{k^2}{m^2} \pm 4 g_{\phi\gamma} k \frac{\phi_0}{m} \cos({\tilde{t}}) \bigg]A_k \, = \, 0 \, .
\ee
As has already been studied in different contexts in \cite{Evan1} (axion monodromy inflation), \cite{Evan2} (inflationary magnetogenesis),\cite{Rudnei} (graviton-induced ALP decay), and \cite{Chunshan} (graviton to photon conversion) this equation has a parametric instability, in this case of tachyonic type.\footnote{The basic mechanisms was already discussed in \cite{JF1, JF2}.}  Infrared modes with
\be
k \, < \, k_c \, \equiv \, 4 g_{\phi \gamma} m \phi_0 
\ee
have a tachyonic mass term for half of the oscillation period of the ultralight dark matter field.  Note that during the half of the oscillation period when the induced effective mass term is positive, the amplitude of $A_k$ oscillates.

To obtain a lower bound on the efficiency of the resonance, we can take the amplitude of $A_k$ to be constant during the half of the oscillation period when the effective mass is positive.  We can subdivide the time interval when the mass term is negative into the interval (one third of the period) when $|\cos(\tilde{t})| > 1/2$ and the rest of the period when the amplitude is smaller.  To get a lower bound on the growth of $A_k$ we take $A_k$ to be constant during the latter sub-interval,  while during the first sub-interval, the amplitude of $A_k$ grows at a faster rate than
\be
A_k({\Delta \tilde{t}}) \, \sim \, e^{ \mu_k {\Delta \tilde{t}} / 2} A_k(0) \, ,
\ee
where the Floquet index $\mu_k$ is given by
\be
\mu(k) \, \sim \, 2 \bigl( g_{\phi \gamma} \phi_0 m^{-1} k  \bigr)^{1/2} \, ,
\ee
and $\Delta {\tilde{t}}$ is the duration of the one-third period.  Averaged over time, we find a lower bound on the amplitude of $A_k$ which is given by
\be
A_k({\tilde{t}}) \, \sim \, e^{ \mu_k {\tilde{t}} / 6} A_k(0) \, ,
\ee
where the extra factor of $1/3$ in the exponent comes from the fact that we take the growth to occur only during a one-third period.

The resonance is efficient on Hubble time scales provided that
\be
\mu(k_c) m t_H \, \gg \, 6\, ,
\ee
where $t_H$ is the Hubble time. Inserting the current Hubble time $t_H \sim 10^{43} {\rm{GeV}}^{-1}$ and the expressions we have derived for $\mu(k)$ and $k_c$
we obtain the condition
\be
{\tilde{g}}_{\phi \gamma} \, \gg \, 10^{-12}
\ee
in order for the resonance to be efficient. A more realistic criterion for efficiency is to use the collapse time $t_h \sim 10^5 {\rm{yrs}} \, \sim \,10^{37} {\rm{GeV}}^{-1}$ in which case the efficiency condition is stricter and yields
\be
{\tilde{g}}_{\phi \gamma} \, \gg \, 10^{- 7} \, .
\ee
In the following we will assume that this condition on the parameters is satisfied.

The next question we wish to answer is whether the parametric excitation of photons is efficient enough to convert a substantial fraction of the halo energy density into photons. For the moment we will neglect back-reaction effects on the resonance process.  Note that the produced photons can back-react on the equation of motion of the ultralight dark matter field and destroy the coherent oscillations which drive the resonance.  Based on energetic considerations, back-reaction will dominate once a fraction $f$ of the order one of the halo density has been converted to photons. We will estimate how long this will take (and set $f = 1$ for simplicity).

Assuming that the photon field starts out in its vacuum state,  
\be
A_k(0) \, = \, \frac{1}{\sqrt{2k}} 
\ee
the photon energy density after a time interval $t$ will be given by
\be
\rho_A(t) \, \sim \, k_c^4 e^{\mu_{k_c} m t / 3} \, .
\ee
This is obtained by integrating over the phase space of modes with $k < k_c$ which undergo resonance, and realizing that the integral is dominated at the upper end $k = k_c$. Inserting the expressions for $\mu_k$ and $k_c$ and making use of (\ref{rel1}) we find that 
\be
\rho_{A} \, = \, \rho_h
\ee
occurs after a time interval of
\be
t \, \sim \, \frac{1}{4 {\tilde{g_{\phi \gamma}}}} 10^{33} {\rm{GeV}}^{-1} 
\ee
which is many orders of magnitude shorter than the Hubble expansion time scale. Thus, we conclude that the resonance process is very efficient at converting ultralight dark matter energy into infrared photons.

At this point, it is important to emphasize that it is only infrared photons which are produced via this parametric resonance instability,  namely only photons with 
\be
k \, < \, k_c \sim {\tilde{g}_{\phi \gamma}} 10^{-21} {\rm{eV} } 
\ee
which is many orders of magnitude lower than the energy of the UV photons required to dissociate $H_2$ in the halo. Thus, in order for our mechanism to be able to realize the DCBH collapse criteria, a process is required to transport photons to smaller wavelengths. 

In the following section, we will review the requirements for successful DCBH formation.  Then, we will study thermalization or energy cascade as a way to satisfy these requirements

\section{Criteria for DCBH Formation}

An important pathway to form SMBHs at high redshifts is via the direct collapse of a pristine gas cloud, which will generate massive black hole seeds with mass $\sim 10^5 M_\odot$ within $\sim 1{\rm{Myr}}$ \cite{DCBH-1,DCBH-2,DCBH-3,DCBH-4,DCBH-5,DCBH-6,DCBH-7,DCBH-8,DCBH-9,DCBH-10,DCBH-11}. In contrast to stellar black hole seeds, DCBHs are able to grow to $10^9M_\odot$ at redshifts $z\gsim 6$ without super-Eddington accretion.

To directly collapse into a black hole, the gas cloud should keep a high enough temperature to undergo a monolithic collapse without fragmentation. Thus, rapid cooling mechanisms, e.g. molecular hydrogen cooling, have to be suppressed in such a halo and we need to restrict the abundance of $H_2$.

In dust-free gas clouds, the main formation mechanism of $H_2$ is a two-step process \cite{H2-formation-1,H2-formation-2}:
\begin{align}
H + e^- &\rightarrow H^- + \gamma \nonumber \\
H + H^- &\rightarrow H_2 + e^-. \nonumber
\end{align}
Thus, strong UV radiation can dissociate $H_2$ via the two processes:
\begin{itemize}
\item[-] photodissociation by Lyman-Werner (LW) radiation (photons in the LW energy band $11.2 {\rm eV} < E < 13.6 {\rm eV}$): $$ H_2 + \gamma_{LW} \rightarrow H+H;$$
\item[-] photodetachment by ultraviolet (UV) photons with energy $0.76 {\rm eV} < E < 13.6 {\rm eV}$: $$H^- + \gamma_{UV} \rightarrow H+e^-.$$
\end{itemize}

Therefore, strong enough UV radiation is a key criterion for DCBH formation. Early works consider a critical flux density of LW radiation $J_{\rm LW,crit}$ \cite{Bryce} 
\begin{align}
J_{\rm LW,crit} \, &\sim \, (10^{-18}-10^{-16}) \,{\rm erg\,s^{-1}cm^{-2} Hz^{-1}Sr^{-1}},\nn
&\sim (10^{-44}-10^{-42}) \,{\rm GeV^3} \label{JLW-required}
\end{align}
and argue that if radiation in the LW energy band exceeds this critical value during monolithic collapse, a DCBH could form in the center of the gas cloud. The large uncertainty in this critical value comes from the shape of the radiation spectrum.

\begin{figure}
  	\includegraphics[width=8cm]{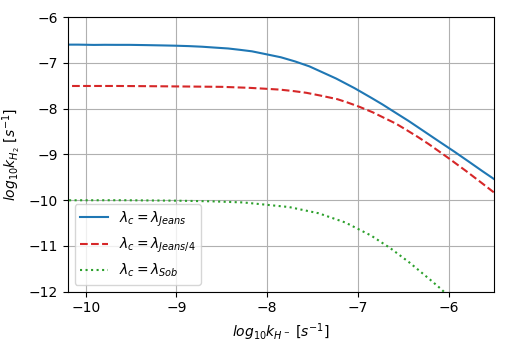}
	\caption{The direct collapse critical curves in the $k_{H^-}-k_{H_2}$  plane from \cite{DCBH-curve-3}. The three curves are determined by 3D hydrodynamic simulations with the self-shielding characteristic length equal to the Jeans length (blue curve), a quarter of the Jeans length (red curve) and the Sobolev-like length (green curve).}
	\label{fig1}
\end{figure}

However, recent works \cite{DCBH-10,DCBH-curve-2} demonstrate that if the photodetachment rate $k_{H^-}$ is sufficiently high, the abundance of molecular hydrogen will be suppressed even if the LW radiation intensity and the photodissociation rate $k_{H_2}$ are not high enough. Hydrodynamic simulations \cite{DCBH-curve-3,DCBH-curve-4} show that there is a direct collapse critical curve in the $k_{H^-}-k_{H_2}$ plane (shown in Fig.\ref{fig1}), which is independent of the radiation spectrum. If the two reaction rates are beyond this curve during the initial stage of collapse,  the $H_2$ fraction remains at a very low level and the evolution of gas temperature is dominated by atomic cooling instead of $H_2$ cooling. 

The photodissociation rate $k_{H_2}$ is approximately determined by the LW radiation intensity \cite{DCBH-curve-4}:
\be
k_{\mathrm{H}_2} \approx 1.39 \times 10^{-12} \mathrm{~s}^{-1}\left(\frac{J_{L W}}{10^{-21} \mathrm{erg} \mathrm{~s}^{-1} \mathrm{~Hz}^{-1} \mathrm{~cm}^{-2} \mathrm{sr}^{-1}}\right) \label{eq-kH2-1}
\ee
where $J_{L W}$ is the average flux in the LW energy range. 

The above equation shows the dissociation rate in a gas cloud that is optically thin for LW radiation. If the column density of molecular hydrogen, $N_{H_2}\equiv n_{H_2}\lambda_c$ (where $\lambda_c$ is the characteristic length of the column), is higher than $\sim 10^{14}cm^{-2}$ during the collapse, the self-shielding effect could play a role and reduce the dissociation rate, yielding
\be
k_{\mathrm{H}_2}\left(N_{\mathrm{H}_2}\right)=k_{\mathrm{H}_2}\left(N_{\mathrm{H}_2}=0\right) f_{s h}\left(N_{\mathrm{H}_2}, T\right) \label{eq-kH2-2}
\ee
where $f_{s h}$ is the shielding fraction.  From numerical simulations, it is a function of the $H_2$ column density and gas cloud temperature \cite{DCBH-curve-2}:
\ba
f_{s h}\left(N_{\mathrm{H}_2}, T\right) \, &=& \, \frac{0.965}{\left(1+x / b_5\right)^{1.1}} + \frac{0.035}{(1+x)^{0.5}} \nonumber \\
& & \, \times \exp \left[-8.5 \times 10^{-4}(1+x)^{0.5}\right]
\ea
where 
\be
x \, \equiv \, N_{\mathrm{H}_2} \times 10^{14} \mathrm{~cm}^{-2} \, ,
\ee
and
\be
b_5 \, \equiv \, b / 10^5 \mathrm{~cm} \mathrm{~s}^{-1} \, ,
\ee
where $b$ is the Doppler broadening parameter.  In Fig. \ref{fig1},  three choices for $\lambda_c$ are considered, namely  $\lambda_c = \lambda_{Jeans}$, $\lambda_c = \lambda_{Jeans}/4$ or $\lambda_c$ given by a Sobolev-type length. Here we focus on the first case to obtain the most conservative result.


Let us now turn to the photodetachment rate which depends on the shape of the radiation spectrum \cite{DCBH-2}:
\be
k_{H^-} \, 
= \int_{0.76 eV}^{13.6 eV} \frac{2dE}{E}\, J(E)\sigma_{H^-}(E), \label{eq-kH-}
\ee
where $\sigma_\nu (cm^2)$ is the photodetachment cross-section \cite{DCBH-curve-2}
\be
\sigma_{\mathrm{H}^{-}}(E) \, \approx \,  4.31 \times 10^{-18} \frac{\hat{E}^3}{\big(0.0555+\hat{E}^2\big)^3} \mathrm{~cm}^2, 
\ee
with
\be 
\hat{E} \, = \, \sqrt{\frac{E-0.754 \mathrm{eV}}{13.6 \mathrm{eV}}}.
\ee

Since the parametric resonance photon production process only generates photons in the far infrared,  a mechanism is required to transfer photons into the UV regime. In the following we will discuss two possible mechanisms: thermalization or (turbulent) energy cascade.  Given any one of these mechanisms, we can then compute the two reaction rates $k_{H2}$ and $k_{H-}$ and study under which conditions the rates are above the critical curve of Fig. \ref{fig1} in the parameter space.

\section{Realizing the DCBH Criteria via Thermalization or Energy Cascade} 

Since the parametric resonance process only produces low--requency photons with $k < k_c \ll k_{LW}$, there needs to be a cascade of photon flux from low to high frequencies. Two obvious mechanisms are thermalization and turbulent cascade.  To start, we compute the initial photon flux before any cascade process.

Assuming that a fraction $f$ of the halo energy is converted into photons via the parametric resonance process discussed in Section III, we can estimate the flux $J$ of photons of the critical energy $k_c$:
\be
J(k_c) \, \sim \, \frac{f}{k_c} \rho_A, \, .
\ee
where $\rho_A$ is the energy density of axions in the halo and could be a function of radius $r$.
 
The strong radiation generated by axions would thermalize the gas cloud to a temperature 
\be
T \, \sim \, \rho_{rad}^{1/4}, \label{ThermalTemp}
\ee 
where $\rho_{rad}$ is the radiation energy density, and will lead to a black body radiation with energy spectrum
\be
J_{black-body}(E,T) = \frac{E^3}{4\pi^3}\frac{1}{e^{E/T}-1}.
\ee
Here we assume that the gas cloud is optically thick for these photons at high redshifts. Note that this spectrum peaks at $E\sim T$, which is usually significantly smaller than the LW energy band, so the corresponding flux is suppressed exponentially and the photodetachment could be crucial.

The second mechanism which we consider is a turbulent cascade in which
\be
J_{cascade}(k) \, \sim \, J^i(k_c) \bigg( \frac{k_c}{k}  \bigg)^n \, ,\label{eq-spec-cascade}
\ee
where $n$ is the scaling index. and the superscript $i$ indicates the pre-turbulent value. For Kolmogorov scaling we have $n = 5/3$ \cite{Kolmogorov}, while for Batchelor scaling $n = 1$ \cite{Batchelor}. 

\subsection{The critical flux density for LW radiation}

Let us first consider the critical flux density for LW radiation in \eqref{JLW-required}.  We will separately study in which cases the thermalization and photon energy cascade mechanisms can lead to a sufficiently high flux of LW radiation.
 
\subsubsection{Thermalization}

Much of the literature on DCBHs adopts a one-zone model, which assumes that the density and temperature are uniform in the gas cloud \cite{DCBH-10,DCBH-curve-2}. However, we will show that thermalization is incapable of generating the required LW flux if the dark matter density profile is not taken into account.
Inserting the virialized density of the halo $\rho_h$ into eq.~\eqref{ThermalTemp}, the thermalized temperature is 
\be
T\sim f^{1/4}10^{-1} {\rm eV} \sim f^{1/4} 10^3 K. \label{eq-thermal-tem}
\ee
Then, we can compute the black-body flux density at $E_{LW} = 13.6eV$
\be
J_{black-body}(E_{LW}) \simeq 2\times10^{-26} \cdot e^{-90 f^{-1/4}} GeV^3
\ee
As the fraction $f$ should be smaller than $1$, the above LW radiation intensity should be smaller than
\be
J_{black-body}(E_{LW}) < 9\times10^{-66} GeV^3,
\ee
which is much smaller than the critical value in eq. \eqref{JLW-required}. Thus,  at first sight, it appears that the black body radiation from a thermalized halo cannot satisfy the DCBH formation criterion.

However,  in the above calculation, we assumed a uniform density in the halo. But note that the density distribution in a halo is not uniform, and taking this non-uniformity into account may lead to a different result. Here we assume the halo follows a Navarro-Frenk-White (NFW) profile \cite{NFW}
\be
\rho(r) = \frac{\rho_{h}}{3\big[\ln(1+c)-\frac{c}{1+c}\big]}\frac{1}{\frac{r}{R_{vir}}\big(c^{-1}+\frac{r}{R_{vir}}\big)^2},
\ee
where $\rho_{h}$ is the average density of the halo in eq. \eqref{eq-halo-dens}, $R_{vir}$ is the virial radius of the halo $R_{vir} \simeq (\frac{3}{4\pi}\frac{M_h}{\rho_h})^{1/3}$, and $c$ is the concentration characterizing the compactness of the halo. Then, the thermalized halo has a temperature as a function of the radius
\be
T_{rad}(r) \simeq 1300 K \cdot \bigg(\frac{f}{A_{NFW}}\frac{1}{\frac{r}{R_{vir}}(c^{-1}+\frac{r}{R_{vir}})^2}\bigg)^{1/4},
\ee
where $A_{NFW} \equiv \big[\ln(1+c)-\frac{c}{1+c}\big]$.

\begin{figure}
  	\includegraphics[width=6cm]{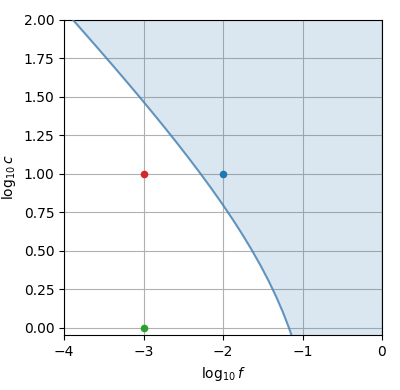}
	\caption{In the shaded region in the $f-c$ plane, the thermalized halo is able to generate strong enough LW radiation.  To obtain this figure we use the value $J_{LW,crit}=10^{-43}GeV^3$ for the critical flux density  The three points in the figure represent three sets of parameter values, for which we compute the corresponding $H_2$ dissociation and detachment rates in Section \ref{Sec4-2} and compare them with the critical curve in Fig.~\ref{fig1} to verify whether they meet the conditions for DCBH formation.}
	\label{fig2}
\end{figure}

We can calculate the average LW radiation by inserting this temperature into the black-body radiation spectrum and averaging it over the volume of the halo.  By comparing the LW flux density with the critical value in eq. \eqref{JLW-required} we obtain an allowed region in the $f-c$ plane, in which the halo is able to generate sufficiently strong LW radiation. This region is shown as the shaded area in Fig.\ref{fig2}, indicating that if the host halo is compact enough and a significant fraction of energy is converted into radiation via parametric resonance, direct collapse could occur. Note that since the LW flux depends exponentially on the temperature, even a small increase in temperature due to enhanced clustering can lead to a sufficiently large LW flux.

\subsubsection{Energy cascade}

In this subsection, we do not assume that the photons produced by the parametric resonance instability thermalize, but that there is an energy cascade (e. g. due to turbulence) which transports power to higher frequencies. We then calculate the LW radiation flux density in the energy cascade scenario and obtain the constraint on the index $n$ by requiring $J^{LW}_{cascade}>J_{LW,crit}$.  The criterion for sufficient LW radiation is
\be
J_{cascade}(k_{LW}) \, \sim \, J(k_c) \bigg( \frac{k_c}{k_{LW}} \bigg)^n \, > \, J_{LW,crit}.
\ee

Let us work out the implications of this condition and consider a halo which virializes at $1+z=20$ (in this case, $t(z=20)\sim 10^8 yr$ so after $\sim 10^6 yr$, the redshift is about $(1+z)_{DCBH}\sim 19.9$, which is enough for the DCBH to grow into a high-redshift SMBH). In this case, the halo density $\rho_h$ is given by (\ref{eq-halo-dens}), and demanding that the ultralight dark matter dominated the halo mass yields the condition (\ref{rel1}) for $ m\phi_0$.
In this case, $k_c$ is given by
\be
k_c\equiv 4 g_{\phi\gamma}m\phi_0 \simeq 1.3\times10^{-20} eV \cdot \tilde{g}_{\phi\gamma}
\ee
and hence $J(k_c)$ is
\be
J(k_c) \sim \frac{f}{k_c}\rho_h \sim \frac{f}{\tilde{g}_{\phi\gamma}}\times 4\times10^{-11}GeV^3 \, .
\ee

\begin{figure}
  	\includegraphics[width=6cm]{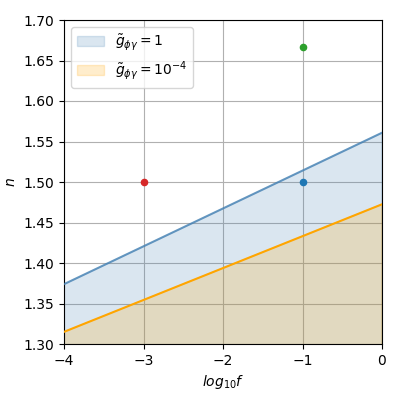}
	\caption{The constraint on the index $n$ of the cascade energy spectrum as a function of the fraction $f$, for $\tilde{g}_{10}=1$ and $\tilde{g}_{10}=10^{-4}$. In the shaded region, a sufficiently high LW background can be generated.}
	\label{fig3}
\end{figure}

For $J_{LW}$
\ba
J(k_{LW}) \, &\sim& \, J(k_c)(\frac{k_c}{k_{LW}})^n \nonumber \\
&=& \,  4\times10^{-11}GeV^3 \cdot \frac{f}{{\tilde{g}}_{\phi\gamma}} \bigg(10^{-21}\tilde{g}_{\phi\gamma}\bigg)^n
\ea
and we require $ J(k_{LW}) > J_{LW,crit} = 2\times10^{-44}GeV^3 $. This leads to the condition
\ba
\frac{f}{\tilde{g}_{\phi\gamma}} \big(10^{-21}\tilde{g}_{\phi\gamma}\big)^n \, &>& 5\times10^{-34}\\
\Rightarrow  n \log_{10}\big(10^{-21}\tilde{g}_{\phi\gamma}\big) \, &>& \,  -34 +\log_{10}\big(5\frac{\tilde{g}_{\phi\gamma}}{f}\big) \nonumber
\ea
which yields
\be
n \, < \, \frac{34+\log_{10}(\tfrac{f}{5\tilde{g}_{\phi\gamma}})}{22-\log_{10}(\tilde{g}_{\phi\gamma})}
\ee
We depict this constraint in Fig.~\ref{fig3} for $\tilde{g}_{\phi\gamma}=1$ and $\tilde{g}_{\phi\gamma}=10^{-4}$.  It depends only weakly on the axion-photon coupling constant $\tilde{g}_{10}$\footnote{Note $\tilde{g}_{\phi\gamma}$ for the orange curve is 4 orders of magnitude smaller than for the blue curve, yet the constraint on the index $n$ decreases by less than 0.1.}, and is always smaller than the Kolmogorov index $n=5/3$. However,  unless the radiation energy fraction $f$ is extremely small, the Batchelor scaling energy spectrum is able to provide the required flux of Lyman-Werner photons.

To obtain an order of magnitude estimate of the results we can take $f = {\tilde{g}}_{\phi \gamma} = 1$ and then we find the condition
\be
n \, < \, \frac{17}{11} \, ,
\ee
which implies that Kolmogorov scaling is not quite sufficient to provide the required flux of Lyman-Werner photons, while Batchelor scaling would be.

Note that in the above we have used the turbulent power index applied to the flux cascade. Had we applied it to energy cascade, then we would need to replace the index $n$ by ${\tilde{n}} + 1$ and we would obtain the more restrictive bound
\be
{\tilde{n}} \, < \, \frac{6}{11} \, .
\ee

As mentioned earlier, the criteria for DCBH formation may be satisfied even if the LW flux is below the critical value used here, as long as the values of the two $H_2$ dissociation rates $k_{H2}$ and $k_{H-}$ in the halo lie above the critical curve in the $k_{H_2}-k_{H^-}$ plane at the initial stage of DCBH formation process. We now turn to a detailed discussion of this point.

\subsection{The critical curve for DCBH formation} \label{Sec4-2}

In this section,  we calculate the photodissociation and photodetachment rates in halos with axion-photon coupling and compare them with the critical curve in Fig.~\ref{fig1} to see under which conditions DCBH formation is possible.  First, we assume that the photons thermalize, and in the second calculation, we assume an energy cascade.

\subsubsection{Thermalization}

\begin{table}
\centering
\caption{The three sets of parameter choices for the thermalization spectrum scenario. The second column indicates their corresponding colors in Figs. \ref{fig2}, \ref{fig4}, and \ref{fig5}. Columns 3 and 4 show the values of the parameters $f$ and $c$ for each set. Columns 5 and 6 show whether the two DCBH formation conditions are satisfied: whether the LW flux density is above the critical value, and whether the values of the reaction rates lie above the critical curve. ``Y'' denotes that the condition is met, and ``N'' indicates that it is not.}
\begin{tabular}{lccccc}
\hline 
~~~~~~~~~  & ~color~ & ~~f~~ & ~~c~~  & $J_{LW,crit}$ & critical curve \\ 
\hline 
Set 1 & blue  & $10^{-2}$ & 10 & Y  & Y \\
Set 2  & red   & $10^{-3}$ & 10 & N  & Y \\
Set 3  & green & $10^{-3}$ & 1  & N  & N \\
\hline
\end{tabular}
\label{table1}
\end{table}

\begin{figure}
  	\includegraphics[width=8cm]{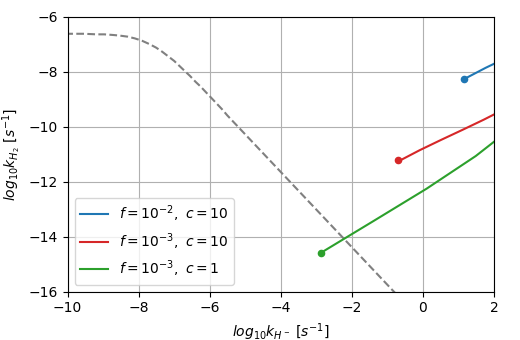}
	\caption{Comparison of the reaction rates in the $k_{H^-}-k_{H_2}$ plane with the direct collapse critical curve, assuming photon thermalization in the halo. The gray dashed line is the highest of the critical curves shown in Fig.~\ref{fig1}, while colored lines represent the photo-detachment and photo-dissociation rates in the thermalized halo. The blue, red and green lines correspond to the parameter sets 1, 2, and 3, respectively. The larger dots on the left end of each line mark the averaged reaction rates for the entire halo with energy density $\rho = \rho_h$, while the rates move to the right along the lines as the enclosed radius $r_0$ decrease.}
	\label{fig4}
\end{figure}

\begin{figure}
  	\includegraphics[width=6cm]{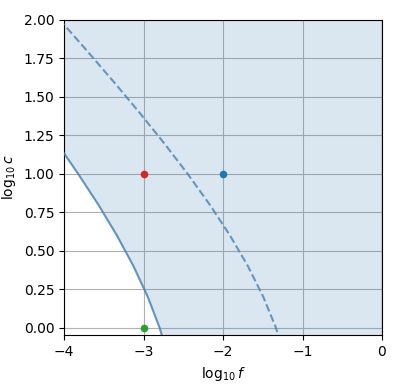}
	\caption{The constraint on the parameters $f$ and $c$ for DCBH formation. The shaded region bounded by the solid line represents the permitted region, in which the thermalized halo is able to emit strong enough UV radiation to keep the reaction rates $k_{H^-}$ and $k_{H_2}$ exceeding the critical curve. The three points denote the parameter values of the three sets summarized in Tab.~\ref{table1}.}
	\label{fig5}
\end{figure}

For each of the two spectra in the above section (thermalized and cascade), we consider three sets of parameter values (the three points in Fig.~\ref{fig2}) and compare the corresponding reaction rates to the direct collapse critical curve. For the thermalized halo with black body radiation spectrum, Tab.~\ref{table1} shows the details of these parameter sets while Fig.~\ref{fig4} illustrates the corresponding reaction rates.

Note that whether a DCBH can form is determined in the initial stages of gas cloud collapse. If the halo does not meet the formation conditions at $\rho\sim\rho_{vir}$, the $H_2$ fraction will increase rapidly and the self-shielding fraction $f_{sh}$ will decrease sharply, which exacerbates the difficulty of dissociating $H_2$. Then, the gas cloud is incapable of collapsing monolithically and forming a DCBH. Therefore, we compare the reaction rate to the critical curve immediately after the virialization and use the halo density in eq.~\eqref{eq-halo-dens}.

We assume an NFW density profile in the thermalization scenario. The stronger self-shielding in the denser central region may affect the realization of the DCBH collapse criteria \cite{DCBH-shielding}. Therefore, in addition to computing the average reaction rates over the entire gas cloud, we also evaluate the rates within a smaller radius to examine how they evolve in the $k_{H_2}-k_{H^-}$ plane. For a radius $r_0<r_h$, we average the flux density $J_{black-body}(E,T(r))$ over $r<r_0$ when computing the reaction rates, and the self-shielding effect is also calculated using the averaged density within $r_0$. 

In Fig.~\ref{fig4}, the coloured lines represent the photo-detachment and photo-dissociation rates for the three parameter sets listed in Tab.~\ref{table1}. The larger dots at the left end of each line mark the reaction rates in the whole halo with the average energy density $\rho = \rho_h$, while the reaction rates move to the right along these lines with the decrease of $r_0$. If the end point of a coloured line is above the critical curve, then the reaction rates $k_{H^-}$ and $k_{H_2}$ are high enough to suppress the abundance of $H_2$ already at the initial stage of the collapse (when the halo is virialized), and hence $H_2$ cooling will be suppressed. Then,  a DCBH could form in such a halo. On the other hand, if the larger dot is below the critical curve (e.g. in the case of the green line with parameter set 3), the fraction of $H_2$ will increase during the collapse and $H_2$ cooling will be important, thus preventing DCBH formation.

Note that, due to the larger value of the ratio $k_{H^-}/k_{H_2}$ in our scenario, the region of interest shifts further to the right in the $k_{H^-}-k_{H_2}$ plane compared to previous works. We extrapolate the critical curve toward the larger $k_{H^-}$ side and assume that the critical curve is still valid in this region.

From this figure, we can see that although the parameter set 2 is outside the permitted region in Fig.~\ref{fig2}, its corresponding reaction rates still exceed the critical curve. This is because the $J_{LW,crit}$ condition mainly considers the photo-dissociation of $H_2$ with the existence of LW radiation. However, a higher photo-detachment rate due to UV radiation with energy lower than $E_{LW}$ will reduce the dependence on $k_{H_2}$, thereby lowering the requirement for the LW radiation flux density.

Furthermore, both of the two reaction rates increase as $r_0$ decreases, indicating that the stronger radiation in the denser region further suppresses $H_2$ formation and the enhancement of the self-shielding effect is insufficient to counteract this influence. Thus, the mean reaction rates in the entire halo provide the most conservative constraints on the parameters and we use them afterwards.

We can numerically compute the constraints on the parameters $f$ and $c$ by requiring that the reaction rates $k_{H^-}$ and $k_{H_2}$ exceed the critical curve. In Fig.~\ref{fig5}, we show the permitted regions corresponding to the direct collapse critical curve (solid curve) and the $J_{LW,crit}$ criterion (dashed curve). The former region is significantly larger, which means that it is easier to reach the critical curve than to get strong enough LW radiation.

Let us summarize these results. For the parameter sets correspond to three typical cases, we obtain:
\begin{itemize}
\item[-] Set 1 satisfies both criteria and in Fig.~\ref{fig4}, its reaction rates are much higher than the critical curve.
\item[-] Set 2 only satisfies the critical curve criterion. Although its photo-dissociation rate is too low, the high photo-detachment rate prevents the formation of $H_2$.
\item[-] Set 3 does not reach the DCBH formation criteria, so $H_2$ will cool the gas cloud to several hundred Kelvin and thus it will fragment during collapse, preventing black hole formation.
\end{itemize}
We use ``Y'' and ``N'' to indicate whether the corresponding parameter set meets the DCBH formation conditions in Tab.~\ref{table1}.

It is straightforward to understand these results: a larger value of $f$ means more radiation,  and thus a higher temperature and more UV radiation; while a larger value of $c$ implies a more compact halo, thus yielding a higher density near the center, which again enhances the amount of radiation. 

\subsubsection{Energy cascade}

In this section, we study the direct collapse critical curve criterion assuming the energy cascade scenario. We numerically calculate the photo-dissociation and photo-detachment rates by inserting the energy cascade spectrum (eq.~\eqref{eq-spec-cascade}) into Eqs.~\eqref{eq-kH2-2} and \eqref{eq-kH-}.

\begin{table}
\centering
\caption{The details of the three parameter sets for the cascade spectrum scenario. The quantities in each column are the same as those in Table 1.}
\begin{tabular}{lccccc}
\hline 
~~~~~~~~~  & ~color~ &   ~~f~~ & ~~n~~ & $J_{LW,crit}$ & critical curve \\ 
\hline
Set 1'  &  blue  &  $0.1$     &  3/2  &  Y  &  Y \\
Set 2'  &  red   &  $10^{-3}$ &  3/2  &  N  &  Y \\
Set 3'  &  green &  $0.1$     &  5/3  &  N  &  N \\
\hline
\end{tabular}
\label{table2}
\end{table}

\begin{figure}
  	\includegraphics[width=8cm]{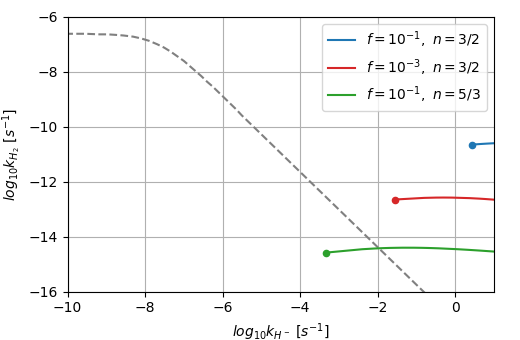}
	\caption{Similar to Fig.~\ref{fig4} but assumiung the energy cascade spectrum. The blue, red, and green lines correspond to the parameter sets 1', 2', and 3',  respectively.}
	\label{fig6}
\end{figure}

\begin{figure}
  	\includegraphics[width=6cm]{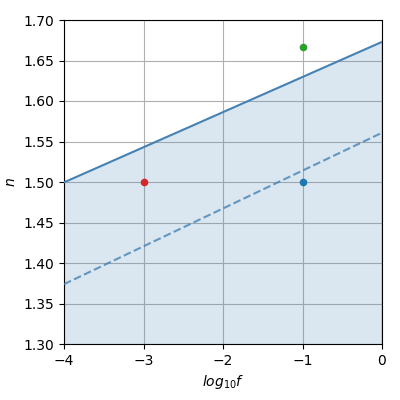}
	\caption{The constraint on the parameters in the energy cascade scenario. In the shaded region of parameter space, the direct collapse critical curve criterion is satisfied.  We only consider the case with coupling constant $\tilde{g}_{10}\equiv 10^{10}g_{\phi\gamma}=1$.}
	\label{fig7}
\end{figure}

Similar to the previous subsection, we consider three parameter sets which correspond to three typical cases,  and their details are listed in Tab.~\ref{table2}. The reaction rates for these parameter sets are shown in Fig.~\ref{fig5}.

From this figure, we can find that in the case of the cascade spectrum the photo-dissociation rate hardly increases as we consider more central regions. This is because, in the cascade scenario, although the intensity of LW radiation slightly increases with the increase of the halo density ($J_{cascade}^{LW}\propto \rho^{(n+1)/2}$), this effect is offset by the enhanced self-shielding effect. In contrast, for the thermalization mechanism, the increase of the halo density will significantly enhance the LW radiation\footnote{The increase in density leads to a rise in the radiation temperature, which shifts the black-body spectrum in the higher energy direction, resulting in an exponential increase in LW radiation as $E_{LW}\gg T$.} and thus, the self-shielding effect is subdominant.

However, as the photodetachment rate significantly increases as $r_0$ decreases, the reaction rates at $\rho=\rho_h$ (corresponding to the larger dots in Fig.~\ref{fig5}) still give rise to the most conservative constraints. 

Figure \ref{fig7} illustrates the permitted regions in the $f-n$ plane where DCBHs can form in the case of the energy cascade scenario. From this figure, we can find that if we consider the Kolmogorov scaling index, i.e. $n = 5/3$, the radiation energy fraction $f$ must be $\sim{\cal O}(1)$ to satisfy the critical curve condition. Conversely, if the Batchelor scaling index ($n=1$) is considered, the constraint on $f$ relaxes to $f\gsim 10^{-15}$.

Whether the three parameter sets in this scenario satisfy the DCBH formation criteria is the same as for the three sets in the thermalization scenario, which are shown in Tab.~\ref{table2}.

\section{Conclusions and Discussion}

We have studied the possibility that photons produced via a parametric resonance instability in the presence of an ultralight dark matter field, which is coherently oscillating in the galactic halo, can lead to sufficient UV radiation to allow a gas cloud accreting onto a seed in the center to reach masses greater than or equal to $10^5 \Msol$ without fragmenting, thus realizing the Direct Collapse Black Hole formation channel.  In order to suppress fragmentation, $H_2$ production must be suppressed, and for this, a sufficiently large amplitude of dissociating flux is required.

Since it is only photons with frequencies in the far infrared which undergo the parametric resonance instability, there needs to be a mechanism that transfers energy to higher-frequency photons in order to obtain sufficient dissociation. We have studied two possibilities: thermalization and energy cascade. In both cases, there are regions in parameter space where the DCBH criteria are met. We compare two criteria which can be used to show that the formation of $H_2$ is suppressed, namely the ``$J_{LW,crit}$ criterion'' and the ``critical curve criterion''. The former considers only photodissociation of $H_2$ and requires sufficient LW flux density, while the latter requires that the photodissociation and photodetachment rates, $k_{H2}$ and $k_{H-}$, lie above the critical curve.  Because $k_{H^-}$ could play an important role in our scenario, the latter criterion shows that $H_2$ formation is suppressed in a wider region of parameter space.

In the case of thermalized photons, the parameter space is parametrized by the halo concentration parameter $c$ and the fraction $f$ of the initial halo dark matter density which is transformed into radiation. In the case of an energy cascade, the parameters are $f$ and $n$, the cascade index. The results are shown in Figs. \ref{fig5} and \ref{fig7}.

We conclude that ultralight dark matter could generate enough photons for DCBH formation via the parametric resonance instability even if the coupling constant between ultralight dark matter and photons is very small. Thus, DCBH formation can occur during the dark ages (as we do not need radiation from stars) and can explain the presence of SMBHs at very high redshifts. 

The mechanism works both in the standard cosmological paradigm of structure formation and also in the presence of cosmic string seeds.  The existence of cosmic string loops could facilitate the formation of DCBHs. First, loops could serve as nonlinear fluctuations in the early universe and could seed halos much earlier than those in $\Lambda$CDM cosmology (soon after $t_{eq}$). Second, halos seeded by loops tend to be significantly more concentrated \cite{JH3}, making it easier to satisfy the DCBH formation criterion in the thermalization scenario.

Note that we have made a number of assumptions in our analysis. First, while the $J_{LW,crit}$ criterion depends on the shape of the radiation spectrum, we used the critical value based on a constant LW radiation background. Secondly, we extrapolated the critical curve beyond the region where has been tested, and we assumed that this extrapolation can be trusted. The validity of these assumptions should be checked.

Note that in a very recent paper \cite{Juerg}, we have used the same coupling studied here,  so show that coherent oscillations of a pseudoscalar dark matter field on cosmological scales can be used to generate magnetic fields on these scales which are sufficiently large to explain the observational lower bounds.

\section*{Acknowledgements}

\noindent  This research is supported in part by funds from NSERC and from the Canada Research Chair program.  JH acknowledges support from a Milton Leung Fellowship in Science. RB wishes to thank Juerg Froehlich for useful discussions, and D. Marsh for communications.



\begin{thebibliography}{99}

\bibitem{Marta}				
M.~Volonteri,  
``Formation of Supermassive Black Holes,''  
Astron.\ Astrophys.\ Rev.\  {\bf 18}, 279 (2010)  
doi:10.1007/s00159-010-0029-x  
[arXiv:1003.4404 [astro-ph.CO]].  



\bibitem{DCBH-1}
V. Bromm and A. Loeb, 
``Formation of the first supermassive black holes'' 
Astrophys.J. \tbf{596} (2003) 34-46, [arXiv:astro-ph/0212400 [astro-ph]]

\bibitem{DCBH-2}
M.G. Haehnelt and M.J. Rees, 
``The formation of nuclei in newly formed galaxies and the evolution of the quasar population'', Mon.Not.Roy.Astron.Soc. \tbf{263} (1993) 168178

\bibitem{DCBH-3}
M. Umemura, A. Loeb, and E.L. Turner, 
``Early cosmic formation of massive black holes'', 
Astrophys.J. \tbf{419} (1993) 459, [arXiv:astro-ph/9303004 [astro-ph]]

\bibitem{DCBH-4}
A. Loeb and F.A. Rasio, 
``Collapse of primordial gas clouds and the formation of quasar black holes'', 
Astrophys.J. \tbf{432} (1994) 52, [arXiv:astro-ph/9401026 [astroph]]

\bibitem{DCBH-5}
D.J. Eisenstein and A. Loeb, 
``Origin of quasar progenitors from the collapse of low spin cosmological perturbations'', 
Astrophys.J. \tbf{443} (1995) 11, [arXiv:astroph/9401016 [astro-ph]]

\bibitem{DCBH-6}
M.C. Begelman, M. Volonteri, and M.J. Rees, 
``Formation of supermassive black holes by direct collapse in pregalactic halos'', 
Mon.Not.Roy.Astron.Soc. \tbf{370} (2006) 289-298, [arXiv:astro-ph/0602363 [astro-ph]]

\bibitem{DCBH-7}
K. Inayoshi, E. Visbal, and Z. Haiman, 
``The Assembly of the First Massive Black Holes'', 
Ann.Rev.Astron.Astrophys. \tbf{58} [arXiv:1911.05791 [astro-ph.GA]]

\bibitem{DCBH-8}
S.P. Oh and Z. Haiman, 
``Second-generation objects in the universe: radiative cooling and collapse of halos with virial temperatures above 104 kelvin'', 
Astrophys.J. \tbf{569} (2002) 558, [arXiv:astro-ph/0108071 [astro-ph]]

\bibitem{DCBH-9}
Z. Haiman, T. Abel, and M.J. Rees, 
``The radiative feedback of the first cosmological objects'', 
Astrophys.J. \tbf{534} (2000) 11-24, [arXiv:astro-ph/9903336 [astro-ph]]

\bibitem{DCBH-10}
Agarwal, Bhaskar, et al. 
``New constraints on direct collapse black hole formation in the early Universe'', 
Mon.Not.Roy.Astron.Soc. \tbf{459} (2016) 4209-4217.

\bibitem{DCBH-11}
Y. Lu, Z. S. C. Picker, and A. Kusenko, 
``Direct collapse supermassive black holes from relic particle decay'', 
Physical Review Letters \tbf{133} (2024) 091001.

\bibitem{Bryce}
B.~Cyr, H.~Jiao and R.~Brandenberger,
``Massive black holes at high redshifts from superconducting cosmic strings,''
Mon. Not. Roy. Astron. Soc. \textbf{517}, no.2, 2221-2230 (2022)
doi:10.1093/mnras/stac1939
[arXiv:2202.01799 [astro-ph.CO]].

\bibitem{CSrevs}				
A.~Vilenkin and E.~P.~S.~Shellard,
  ``Cosmic Strings and Other Topological Defects,''
  (Cambridge Univ. Press, Cambridge, 2000);\\
M.~B.~Hindmarsh and T.~W.~B.~Kibble,
  ``Cosmic strings,''
  Rept.\ Prog.\ Phys.\  {\bf 58}, 477 (1995)
  doi:10.1088/0034-4885/58/5/001
  [hep-ph/9411342];\\
R.~H.~Brandenberger,
  ``Topological defects and structure formation,''
  Int.\ J.\ Mod.\ Phys.\ A {\bf 9}, 2117 (1994)
  doi:10.1142/S0217751X9400090X
  [astro-ph/9310041].


\bibitem{JH1}
H.~Jiao, R.~Brandenberger and A.~Refregier,
``Early structure formation from cosmic string loops in light of early JWST observations,''
Phys. Rev. D \textbf{108} (2023) no.4, 043510
doi:10.1103/PhysRevD.108.043510
[arXiv:2304.06429 [astro-ph.CO]].

\bibitem{JH2}
H.~Jiao, R.~Brandenberger and A.~Refregier,
``N-body simulation of early structure formation from cosmic string loops,''
Phys. Rev. D \textbf{109}, no.12, 123524 (2024)
doi:10.1103/PhysRevD.109.123524
[arXiv:2402.06235 [astro-ph.CO]].

\bibitem{JH3}
S.~M.~Koehler, H.~Jiao and R.~Kannan,
``Investigating cosmic strings using large-volume hydrodynamical simulations in the context of JWST's massive UV-bright galaxies,''
[arXiv:2412.00182 [astro-ph.CO]].

\bibitem{Bram}				
S.~F.~Bramberger, R.~H.~Brandenberger, P.~Jreidini and J.~Quintin, 
``Cosmic String Loops as the Seeds of Super-Massive Black Holes,'' 
JCAP {\bf 1506}, no. 06, 007 (2015) 
doi:10.1088/1475-7516/2015/06/007 
[arXiv:1503.02317 [astro-ph.CO]].

\bibitem{IMBH}
R.~Brandenberger, B.~Cyr and H.~Jiao,
``Intermediate mass black hole seeds from cosmic string loops,''
Phys. Rev. D \textbf{104}, no.12, 123501 (2021)
doi:10.1103/PhysRevD.104.123501
[arXiv:2103.14057 [astro-ph.CO]].

\bibitem{Kusenko}
Y.~Lu, Z.~S.~C.~Picker and A.~Kusenko,
``Direct Collapse Supermassive Black Holes from Relic Particle Decay,''
Phys. Rev. Lett. \textbf{133}, no.9, 091001 (2024)
doi:10.1103/PhysRevLett.133.091001
[arXiv:2404.03909 [astro-ph.GA]].

\bibitem{Elisa}
E.~G.~M.~Ferreira,
``Ultra-light dark matter,''
Astron. Astrophys. Rev. \textbf{29}, no.1, 7 (2021)
doi:10.1007/s00159-021-00135-6
[arXiv:2005.03254 [astro-ph.CO]].

\bibitem{Hui}
L.~Hui,
``Wave Dark Matter,''
Ann. Rev. Astron. Astrophys. \textbf{59}, 247-289 (2021)
doi:10.1146/annurev-astro-120920-010024
[arXiv:2101.11735 [astro-ph.CO]].

\bibitem{DK}
A.~D.~Dolgov and D.~P.~Kirilova,
  ``On Particle Creation By A Time Dependent Scalar Field,''
  Sov.\ J.\ Nucl.\ Phys.\  {\bf 51}, 172 (1990)
  [Yad.\ Fiz.\  {\bf 51}, 273 (1990)].
  
\bibitem{TB}
J.~H.~Traschen and R.~H.~Brandenberger,
  ``Particle Production During Out-of-equilibrium Phase Transitions,''
  Phys.\ Rev.\ D {\bf 42}, 2491 (1990).
  doi:10.1103/PhysRevD.42.2491
  
\bibitem{KLS1}
L.~Kofman, A.~D.~Linde and A.~A.~Starobinsky,
``Reheating after inflation,''
Phys. Rev. Lett. \textbf{73}, 3195-3198 (1994)
doi:10.1103/PhysRevLett.73.3195
[arXiv:hep-th/9405187 [hep-th]].

\bibitem{STB}
Y.~Shtanov, J.~H.~Traschen and R.~H.~Brandenberger,
``Universe reheating after inflation,''
Phys. Rev. D \textbf{51}, 5438-5455 (1995)
doi:10.1103/PhysRevD.51.5438
[arXiv:hep-ph/9407247 [hep-ph]].

\bibitem{KLS2}
L.~Kofman, A.~D.~Linde and A.~A.~Starobinsky,
``Towards the theory of reheating after inflation,''
Phys. Rev. D \textbf{56}, 3258-3295 (1997)
doi:10.1103/PhysRevD.56.3258
[arXiv:hep-ph/9704452 [hep-ph]].

\bibitem{RHrevs}
R.~Allahverdi, R.~Brandenberger, F.~Y.~Cyr-Racine and A.~Mazumdar,
  ``Reheating in Inflationary Cosmology: Theory and Applications,''
  Ann.\ Rev.\ Nucl.\ Part.\ Sci.\  {\bf 60}, 27 (2010)
  doi:10.1146/annurev.nucl.012809.104511
  [arXiv:1001.2600 [hep-th]];\\
M.~A.~Amin, M.~P.~Hertzberg, D.~I.~Kaiser and J.~Karouby,
  ``Nonperturbative Dynamics Of Reheating After Inflation: A Review,''
  Int.\ J.\ Mod.\ Phys.\ D {\bf 24}, 1530003 (2014)
  doi:10.1142/S0218271815300037
  [arXiv:1410.3808 [hep-ph]].

\bibitem{lower}
K.~K.~Rogers and H.~V.~Peiris,
``Strong Bound on Canonical Ultralight Axion Dark Matter from the Lyman-Alpha Forest,''
Phys. Rev. Lett. \textbf{126}, no.7, 071302 (2021)
doi:10.1103/PhysRevLett.126.071302
[arXiv:2007.12705 [astro-ph.CO]].

\bibitem{Rodd}
D.~Y.~Cheong, N.~L.~Rodd and L.~T.~Wang,
``A Quantum Description of Wave Dark Matter,''
[arXiv:2408.04696 [hep-ph]].

\bibitem{AxionLimits}
Ciaran O'Hare, ``cajohare/AxionLimits: AxionLimits,'' https://cajohare.github.io/AxionLimits/.


\bibitem{constraint}
D.~J.~E.~Marsh,
``Axion Cosmology,''
Phys. Rept. \textbf{643}, 1-79 (2016)
doi:10.1016/j.physrep.2016.06.005
[arXiv:1510.07633 [astro-ph.CO]].

\bibitem{JF1}
J.~Frohlich and B.~Pedrini,
``New applications of the chiral anomaly,''
[arXiv:hep-th/0002195 [hep-th]].

\bibitem{JF2}
J.~Frohlich and B.~Pedrini,
``Axions, quantum mechanical pumping, and primeval magnetic fields,''
[arXiv:cond-mat/0201236 [cond-mat]].

\bibitem{Joyce}
M.~Joyce and M.~E.~Shaposhnikov,
``Primordial magnetic fields, right-handed electrons, and the Abelian anomaly,''
Phys. Rev. Lett. \textbf{79}, 1193-1196 (1997)
doi:10.1103/PhysRevLett.79.1193
[arXiv:astro-ph/9703005 [astro-ph]].

\bibitem{Juerg}
R.~Brandenberger, J.~Fr\"ohlich and H.~Jiao,
``Cosmological Magnetic Fields from Ultralight Dark Matter,''
[arXiv:2502.19310 [hep-ph]].

\bibitem{Zeld}
Y.~B.~Zeldovich,
``Gravitational instability: An Approximate theory for large density perturbations,''
Astron. Astrophys. \textbf{5} (1970), 84-89.

\bibitem{Evan1}
E.~McDonough, H.~Bazrafshan Moghaddam and R.~H.~Brandenberger,
``Preheating and Entropy Perturbations in Axion Monodromy Inflation,''
JCAP \textbf{05}, 012 (2016)
doi:10.1088/1475-7516/2016/05/012
[arXiv:1601.07749 [hep-th]].

\bibitem{Evan2}
H.~Bazrafshan Moghaddam, E.~McDonough, R.~Namba and R.~H.~Brandenberger,
``Inflationary magneto-(non)genesis, increasing kinetic couplings, and the strong coupling problem,''
Class. Quant. Grav. \textbf{35}, no.10, 105015 (2018)
doi:10.1088/1361-6382/aaba22
[arXiv:1707.05820 [astro-ph.CO]].

\bibitem{Rudnei}
R.~Brandenberger, V.~Kamali and R.~O.~Ramos,
``Decay of ALP condensates via gravitation-induced resonance,''
JCAP \textbf{11}, 009 (2023)
doi:10.1088/1475-7516/2023/11/009
[arXiv:2303.14800 [hep-ph]].

\bibitem{Chunshan}
R.~Brandenberger, P.~C.~M.~Delgado, A.~Ganz and C.~Lin,
``Graviton to photon conversion via parametric resonance,''
Phys. Dark Univ. \textbf{40}, 101202 (2023)
doi:10.1016/j.dark.2023.101202
[arXiv:2205.08767 [gr-qc]].

\bibitem{H2-formation-1}
P.~C.~Stancil and A.~Dalgarno, 
``Chemical processes in astrophysical radiation fields'', 
Faraday Discussions 109, 61 (1998).

\bibitem{H2-formation-2}
S.~Lepp, P.~C.~Stancil, and A.~Dalgarno, 
``TOPICAL REVIEW: Atomic and molecular processes in the early Universe'', 
Journal of Physics B Atomic Molecular Physics 35, R57-R80 (2002).



\bibitem{DCBH-curve-2}
J. Wolcott-Green, Z. Haiman, and G. L. Bryan, 
``Beyond Jcrit: a critical curve for suppression of H2-cooling in protogalaxies,'' 
Mon. Not. Roy. Astron. Soc. 469, 3329-3336 (2017), 
[arXiv:1609.02142 [astro-ph.GA]].

\bibitem{DCBH-curve-3}
Y. Luo, I. Shlosman, K. Nagamine, and T. Fang, 
``Direct collapse to supermassive black hole seeds: the critical conditions for suppression of $H_2$ cooling,'' 
Mon. Not. Roy. Astron. Soc 492, 4917-4926 (2020).

\bibitem{DCBH-curve-4}
A. Friedlander, S. Schon, and A. C. Vincent,
``Supermassive black hole seeds from sub-keV dark matter,'' 
JCAP, 2023, 2023(06): 033, [arXiv:2212.11100v2 [hep-ph]].

\bibitem{Kolmogorov}
A. N. Kolmogorov,
``A refinement of previous hypotheses concerning the local structure of turbulence in a viscous incompressible fluid at high Reynolds number''. 
Journal of Fluid Mechanics. 13 (1), 82 (1961)
doi:10.1017/S0022112062000518.

\bibitem{Batchelor}
G.K.  Batchelor, 
``Small-scale variation of convected quantities like temperature in turbulent fluid. Part 1. General discussion and the case of small conductivity'', 
Journal of Fluid Mechanics, 5,  113,  (1959), 
Bibcode:1959JFM.....5..113B, doi:10.1017/s002211205900009x, S2CID 122304345

\bibitem{NFW}
J.~F.~Navarro, C.~S.~Frenk and S.~D.~M.~White,
``A Universal density profile from hierarchical clustering,''
Astrophys. J. \textbf{490}, 493-508 (1997)
doi:10.1086/304888
[arXiv:astro-ph/9611107 [astro-ph]].

\bibitem{DCBH-shielding}
Sullivan1, Z.~Haiman1, M.~Kulkarni, and E.~Visbal,
``Can supermassive stars form in protogalaxies due to internal Lyman-Werner feedback?'',
[arXiv:2501.12986[astro-ph.GA]]


\end{thebibliography}
\end{document}